\documentclass[fleqn,usenatbib]{mnras}

\usepackage{newtxtext,newtxmath}

\usepackage[T1]{fontenc}
\usepackage{ae,aecompl}
\usepackage{lscape}

\usepackage{graphicx}	
\usepackage{amsmath}	
\usepackage{amssymb}	

\usepackage[utf8]{inputenc}
\usepackage{hyperref}
\usepackage[export]{adjustbox}
\usepackage{natbib}
\usepackage{xspace}

\newcommand{\msun}{$M_{\odot}$\xspace}
\newcommand{\rsun}{$R_{\odot}$\xspace}
\newcommand{\mearth}{$M_{\earth}$\xspace}
\newcommand{\rearth}{$R_{\earth}$\xspace}
\newcommand{\mstar}{\ensuremath{M_{\star}}\xspace}
\newcommand{\rstar}{\ensuremath{R_{\star}}\xspace}

\newcommand{\feh}{\ensuremath{[\mbox{Fe}/\mbox{H}]}\xspace}
\newcommand{\teff}{\ensuremath{T_{\mathrm{eff}}}\xspace}
\newcommand{\logg}{\ensuremath{\log g}\xspace}
\newcommand{\vsini}{\ensuremath{v \sin i}\xspace}
\newcommand{\gcc}{g\,cm$^{-3}$\xspace}
\newcommand{\rhostar}{\ensuremath{\rho_{\star}}\xspace}
\newcommand{\rp}{\ensuremath{R_\mathrm{p}}\xspace}
\newcommand{\teq}{$T_{\mathrm{eq}}$\xspace}

\newcommand{\kepler}{\textit{Kepler}\xspace}
\newcommand{\ktwo}{\textit{K2}\xspace}
\newcommand{\spitzer}{\textit{Spitzer}\xspace}

\newcommand{\jwst}{\textit{JWST}\xspace}

\newcommand{\wise}{\textit{WISE}\xspace}

\newcommand{\target}{K2-264\xspace}
\newcommand{\epic}{EPIC\,211964830\xspace}
\newcommand{\isochrones}{{\tt isochrones}\xspace}
\newcommand{\vespa}{{\tt vespa}\xspace}

\title[\target]{\target: A transiting multi-planet system in the Praesepe open cluster}

\author[J. H. Livingston et al.]{
John H. Livingston,$^{1,2}$\thanks{E-mail: livingston@astron.s.u-tokyo.ac.jp}
Fei Dai,$^{3,4}$
Teruyuki Hirano,$^{5}$
Davide Gandolfi,$^{6}$
\newauthor
Alessandro A. Trani,$^{1,2}$
Grzegorz Nowak,$^{7,8}$
William D. Cochran,$^{9}$
Michael Endl,$^{9}$
\newauthor
Simon Albrecht,$^{10}$
Oscar Barragan,$^{6}$
Juan Cabrera,$^{11}$
Szilard Csizmadia,$^{11}$
\newauthor
Jerome P. de Leon,$^{1}$
Hans Deeg,$^{7,8}$
Philipp Eigm\"uller,$^{11,19}$
Anders Erikson,$^{11}$
\newauthor
Malcolm Fridlund,$^{12,13}$
Akihiko Fukui,$^{7,14}$
Sascha Grziwa,$^{15}$
Eike W. Guenther,$^{16}$
\newauthor
Artie P. Hatzes,$^{16}$
Judith Korth,$^{15}$
Masayuki Kuzuhara,$^{17,18}$
Pilar Monta{\~n}es,$^{7,8}$
\newauthor
Norio Narita,$^{1,7,17,18}$
David Nespral,$^{7,8}$
Enric Palle,$^{7,8}$
Martin P\"atzold,$^{15}$
\newauthor
Carina M. Persson,$^{13}$
Jorge Prieto-Arranz,$^{7,8}$
Heike Rauer,$^{11,19,20}$
\newauthor
Motohide Tamura,$^{1,17,18}$
Vincent Van Eylen,$^{12}$
Joshua N. Winn$^{4}$
\\
$^{1}$Department of Astronomy, University of Tokyo, 7-3-1 Hongo, Bunkyo-ky, Tokyo 113-0033, Japan\\
$^{2}$JSPS Fellow\\
$^{3}$Dept. of Physics and Kavli Institute for Astrophysics and Space Research, Massachusetts Institute of Technology, Cambridge, MA, 02139, USA\\
$^{4}$Department of Astrophysical Sciences, Princeton University, 4 Ivy Lane, Princeton, NJ 08544, USA\\
$^{5}$Department of Earth and Planetary Sciences, Tokyo Institute of Technology, 2-12-1 Ookayama, Meguro-ku, Tokyo 152-8551, Japan\\
$^{6}$Dipartimento di Fisica, Universit\`a di Torino, via P. Giuria 1, 10125 Torino, Italy\\
$^{7}$Instituto de Astrof\'\i sica de Canarias, C/\,V\'\i a L\'actea s/n, 38205 La Laguna, Spain\\
$^{8}$Departamento de Astrof\'isica, Universidad de La Laguna, 38206 La Laguna, Spain\\
$^{9}$Department of Astronomy and McDonald Observatory, University of Texas at Austin, 2515 Speedway,~Stop~C1400,~Austin,~TX~78712,~USA\\
$^{10}$Stellar Astrophysics Centre, Department of Physics and Astronomy, Aarhus University, Ny Munkegade 120, DK-8000 Aarhus C, Denmark\\
$^{11}$Institute of Planetary Research, German Aerospace Center, Rutherfordstrasse 2, 12489 Berlin, Germany\\
$^{12}$Leiden Observatory, Leiden University, 2333CA Leiden, The Netherlands\\
$^{13}$Department of Space, Earth and Environment, Chalmers University of Technology, Onsala Space Observatory, 439 92 Onsala, Sweden\\
$^{14}$Subaru Telescope Okayama Branch Off., Nat. Astronom. Obs. of Japan, NINS, 3037-5 Honjo, Kamogata, Asakuchi, Okayama 719-0232, Japan\\
$^{15}$Rheinisches Institut f\"ur Umweltforschung an der Universit\"at zu K\"oln, Aachener Strasse 209, 50931 K\"oln, Germany\\
$^{16}$Th\"uringer Landessternwarte Tautenburg, Sternwarte 5, D-07778 Tautenberg, Germany\\
$^{17}$Astrobiology Center, NINS, 2-21-1 Osawa, Mitaka, Tokyo 181-8588, Japan\\
$^{18}$National Astronomical Observatory of Japan, NINS, 2-21-1 Osawa, Mitaka, Tokyo 181-8588, Japan\\
$^{19}$Center for Astronomy and Astrophysics, TU Berlin, Hardenbergstr. 36, 10623 Berlin, Germany\\
$^{20}$Institute of Geological Sciences, FU Berlin, Malteserstr. 74-100, D-12249 Berlin, Germany
}

\date{Accepted XXX. Received YYY; in original form ZZZ}

\pubyear{2018}

\begin{document}
\label{firstpage}
\pagerange{\pageref{firstpage}--\pageref{lastpage}}
\maketitle

\begin{abstract}
Planet host stars with well-constrained ages provide a rare window to the time domain of planet formation and evolution. The NASA \ktwo mission has enabled the discovery of the vast majority of known planets transiting stars in clusters, providing a valuable sample of planets with known ages and radii. We present the discovery of two planets transiting \target, an M2 dwarf in the intermediate age (600-800 Myr) Praesepe open cluster (also known as the Beehive Cluster, M44, or NGC 2632), which was observed by \ktwo during Campaign 16. The planets have orbital periods of 5.8 and 19.7 days, and radii of $2.2 \pm 0.2 $ and $2.7 \pm 0.2$ \rearth, respectively, and their equilibrium temperatures are $496 \pm 10$ and $331 \pm 7$ K, making this a system of two warm sub-Neptunes. When placed in the context of known planets orbiting field stars of similar mass to \target, these planets do not appear to have significantly inflated radii, as has previously been noted for some cluster planets. As the second known system of multiple planets transiting a star in a cluster, \target should be valuable for testing theories of photoevaporation in systems of multiple planets. Follow-up observations with current near-infrared (NIR) spectrographs could yield planet mass measurements, which would provide information about the mean densities and compositions of small planets soon after photoevaporation is expected to have finished. Follow-up NIR transit observations using \spitzer or large ground-based telescopes could yield improved radius estimates, further enhancing the characterization of these interesting planets.
\end{abstract}




\section{Introduction}

The great wealth of data from large exoplanet surveys is a powerful tool for statistical studies of planet formation and evolution. For example, the large number of transiting planets, mostly discovered by the \kepler mission, has enabled the discovery of detailed structure in the observed planetary radius distribution \citep{2017AJ....154..109F, 2018MNRAS.479.4786V, 2018arXiv180501453F, 2018arXiv180500231B}, which had been predicted by theories of planetary evolution via photoevaporation \citep[e.g.][]{2013ApJ...775..105O, 2014ApJ...792....1L}. The observed properties of planets are intrinsically dependent on the properties of their host stars; indeed, the phrase ``know thy star, know thy planet'' has become ubiquitous in the field of exoplanet science.

Besides the necessity of host star characterization for obtaining planet properties from indirect measurements, the comparison of planet properties with those of their hosts has long been a source of great interest \citep[e.g.][]{2005ApJ...622.1102F, 2018AJ....155...89P}, as the discovery of a causal relationship would provide a rare glimpse of the mechanisms underpinning planet formation and the processes sculpting them thereafter. However, the vast majority of known planet host stars are of uncertain age, so planet demographics and occurrence rates have been largely unexplored in the time domain. Planets orbiting stars in clusters thus present a rare opportunity for investigations of planet properties as a function of time.

Most of the first known planets orbiting cluster stars were discovered by the radial velocity (RV) method \citep[e.g.][]{2007ApJ...661..527S, 2007A&A...472..657L, 2012ApJ...756L..33Q, 2014ApJ...787...27Q, 2016A&A...588A.118M}. However, an inherent limitation of the RV method is that most planets discovered in this way do not transit their host stars, so their radii are unknown and the measured masses are lower limits. By extending the \kepler mission to the ecliptic plane, the \ktwo mission \citep{2014PASP..126..398H} has enabled the discovery of the vast majority of transiting planets in clusters \citep{2016AJ....152..223O,2017AJ....153..177P,2016AJ....151..112D,2016ApJ...818...46M, 2017AJ....153...64M,2017MNRAS.464..850G,2018AJ....155...10C,2018AJ....155....4M,2018AJ....155..115L}, including the youngest known transiting planet \citep{2016Natur.534..658D,2016AJ....152...61M}.

We present here the discovery of two planets transiting \target, a low mass star in the Praesepe open cluster. We identified two sets of transits in the \ktwo photometric data collected during Campaign 16, then obtained high resolution adaptive optics (AO) imaging of the host star. Precise photometry and astrometry from the {\it Gaia} mission \citep{2016A&A...595A...1G}, along with archival data, enable the characterization of the host star and facilitate the interpretation of the transit signals. We combine the results of detailed light curve analyses and host star characterization to determine the planetary nature of the transit signals, as well as constrain fundamental properties of the two small planets. \target is now the second known transiting multi-planet system in a cluster, offering a rare glimpse into the time domain of planet formation and evolution; its discovery thus significantly enhances a crucial avenue for testing theories of migration and photoevaporation. The transit detections and follow-up observations that enabled this discovery are the result of an international collaboration called KESPRINT. While this manuscript was in preparation \citet{2018arXiv180807068R} announced an independent discovery of this system. Given the rarity of transiting multi-planet cluster systems, it is not surprising that multiple teams pursued follow-up observations of this valuable target.

This paper is organized as follows. In \autoref{sec:obs}, we describe the \ktwo photometry and high resolution imaging of the host star, as well as archival data used in our analysis. In \autoref{sec:analysis} we describe our transit analyses, host star characterization, planet validation, and dynamical analyses of the system. Finally, we discuss the properties of the planet system and prospects for future studies in \autoref{sec:discuss}, and we conclude with a summary in \autoref{sec:summary}.

\section{Observations}
\label{sec:obs}

\subsection{K2 photometry}

\begin{figure}
	\centering
	\includegraphics[clip,trim={0cm 0cm 0cm 0cm},width=0.5\textwidth]{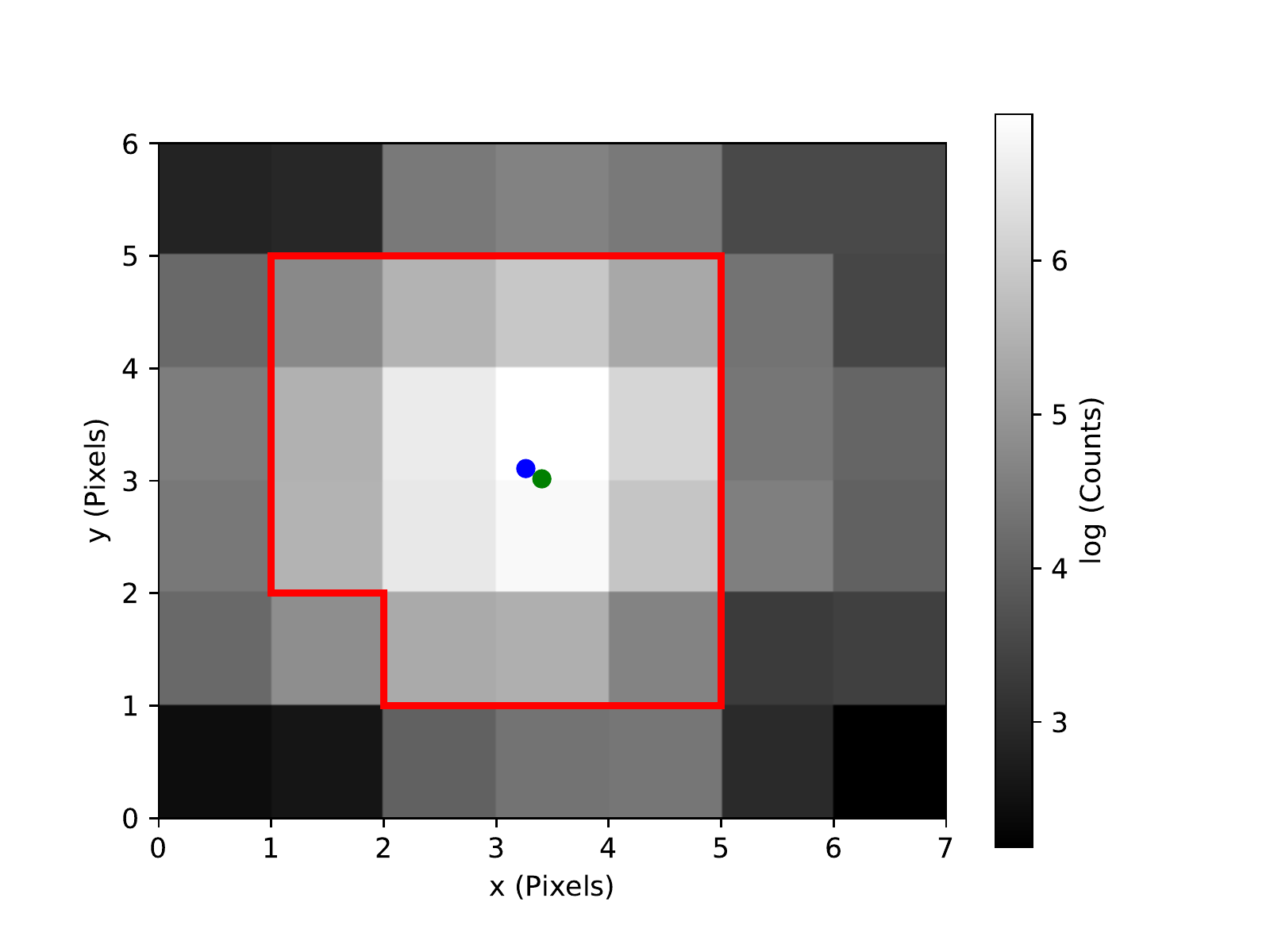}
	\caption{\ktwo ``postage stamp'' of \target with a 1.8 pixel ($\sim$7\arcsec) photometric aperture overplotted in red. The green circle indicates the current position of the target in the EPIC, and the blue circle is the center of the flux distribution.}
	\label{fig:aper}
\end{figure}

\begin{figure*}
	\centering
	\includegraphics[clip,trim={0cm 0cm 1cm 0cm},width=\textwidth]{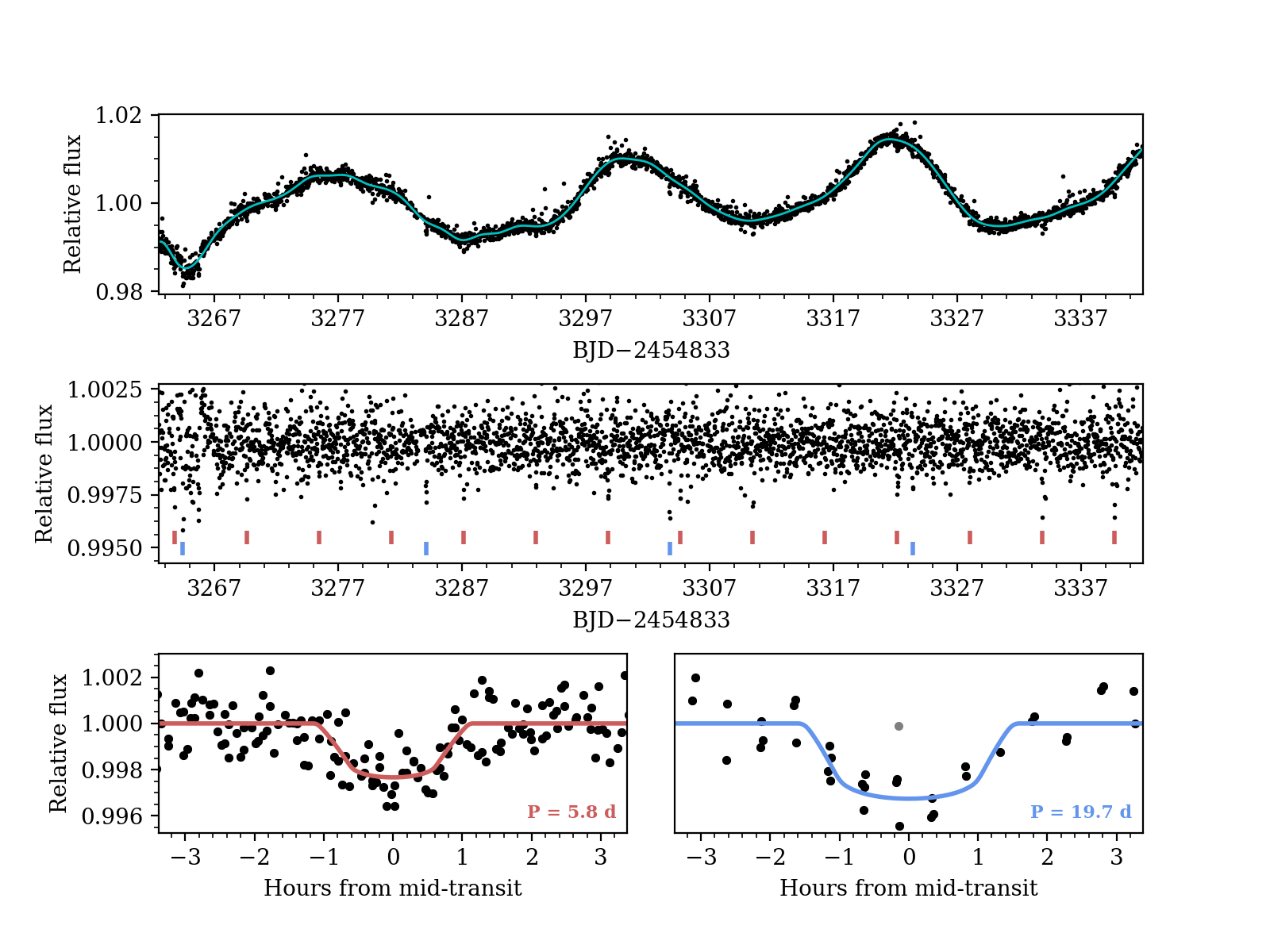}
	\caption{\ktwo photometry of \target in black with cubic spline fit overplotted in cyan (top), flattened light curve with the transits of the planets indicated by tick marks (middle), and the same photometry phase-folded on the orbital period of each planet (bottom). The best-fitting transit models are shown in red and blue for planets b and c, respectively, which also correspond to the color of the tick marks in the top panel.}
	\label{fig:k2}
\end{figure*}

\target (also known as \epic, Cl* NGC 2632 JS 597, 2MASS J08452605+1941544, and {\it Gaia} DR2 661167785238757376) was one of 35,643 long cadence (LC) targets observed during Campaign 16 of the \ktwo mission, from 2017-12-07 23:01:18 to 2018-02-25 12:39:52 UT. \target was proposed as a LC target by GO programs 16022 (PI Rebull), 16031 (PI Endl), 16052 (PI Stello), and 16060 (PI Agueros). The data were downlinked from the spacecraft and subsequently calibrated and made available on the Mikulski Archive for Space Telescopes\footnote{\url{https://archive.stsci.edu/k2/}} (MAST). We describe our light curve preparation and transit search procedures in detail in \citet{2018AJ....156...78L}. In brief, we extracted photometry from the \ktwo pixel data with circular apertures and applied a correction for the systematic caused by the pointing drift of \ktwo, similar to the approach described by \citet{2014PASP..126..948V}. The apertures did not use partial pixels, so a given pixel was included if its center was within the aperture radius. For a range of aperture radii up to four pixels, we computed the 6-hour combined differential photometric precision \citep[CDPP;][]{2012PASP..124.1279C} of the resulting light curve. The light curves did not exhibit any significant variation of transit depth with aperture size. For \target, we selected an aperture with a 1.8 pixel radius (see \autoref{fig:aper}), as this resulted in the corrected light curve with the lowest CDPP value. We then removed stellar variability using a cubic spline with knots every 1.5 days, and searched the light curve for transits using the Box-Least-Squares algorithm \citep[BLS,][]{2002A&A...391..369K}. We identified two candidate planets with signal detection efficiency \citep{2014A&A...561A.138O} values of 11.0 and 10.1. The light curve and phase-folded transits of \target are shown in \autoref{fig:k2}. Subsequent modeling described in \autoref{sec:transit} yielded transit SNR \citep{2018AJ....156...78L} values of 14.0 and 15.9 for the inner and outer planet candidates, respectively. We identified an outlier most likely caused by residual systematics in the light curve and excluded it from our transit analysis (see gray data point in lower right panel of \autoref{fig:k2}).

\subsection{Subaru/IRCS adaptive optics imaging}

\begin{figure}
	\includegraphics[clip,trim={0cm 0cm 0cm 0cm},width=\columnwidth]{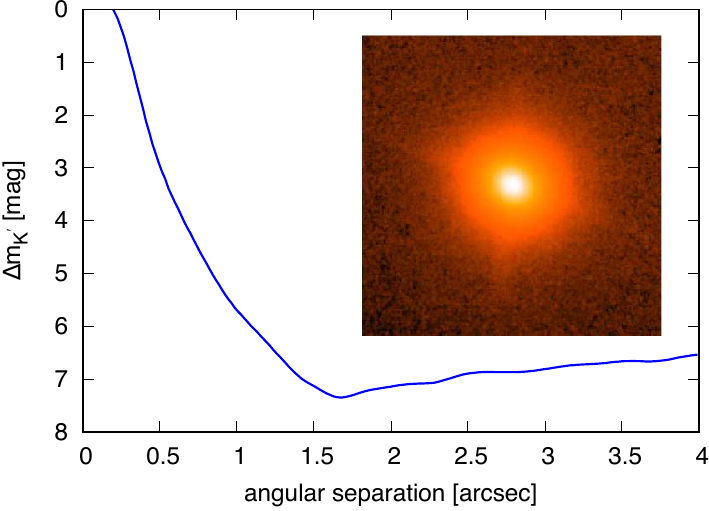}
	\caption{5-$\sigma$ background sensitivity limit (blue curve) and inset 4\arcsec$\times$4\arcsec\, image of \target (inset). The x-axis is angular separation from \target in arcseconds, and the y-axis is differential magnitude in the $K_s$ band.}
	\label{fig:ao}
\end{figure}

On UT 2018 June 14, we obtained high resolution adaptive optics (AO) imaging of \target with the IRCS instrument mounted on the 8.2 meter Subaru telescope on Mauna Kea, HI, USA. The AO imaging utilized the target stars themselves as natural guide stars. We adopted the fine sampling mode ($1\,\mathrm{pix}\approx 20\,\mathrm{mas}$) and five-point dithering, and a total exposure time of 300 seconds was spent for \target. The full width at the half maximum (FWHM) of the target image was $\sim 0\farcs 22$ after the AO correction. Following \citet{2016ApJ...820...41H}, we performed dark current subtraction, flat fielding, and distortion correction before finally aligning and median combining the individual frames. In this manner we produced and visually inspected 16\arcsec$\times$16\arcsec\, combined image, which we then used to compute a 5-$\sigma$ contrast curve following the procedure described in \citet{2018AJ....155..127H}. We show the resulting contrast curve with a 4\arcsec$\times$4\arcsec\, image of \target inset in \autoref{fig:ao}.

\subsection{Archival imaging}

\begin{figure*}
	\includegraphics[clip,trim={0cm 0cm 0cm 0cm},width=0.8\textwidth]{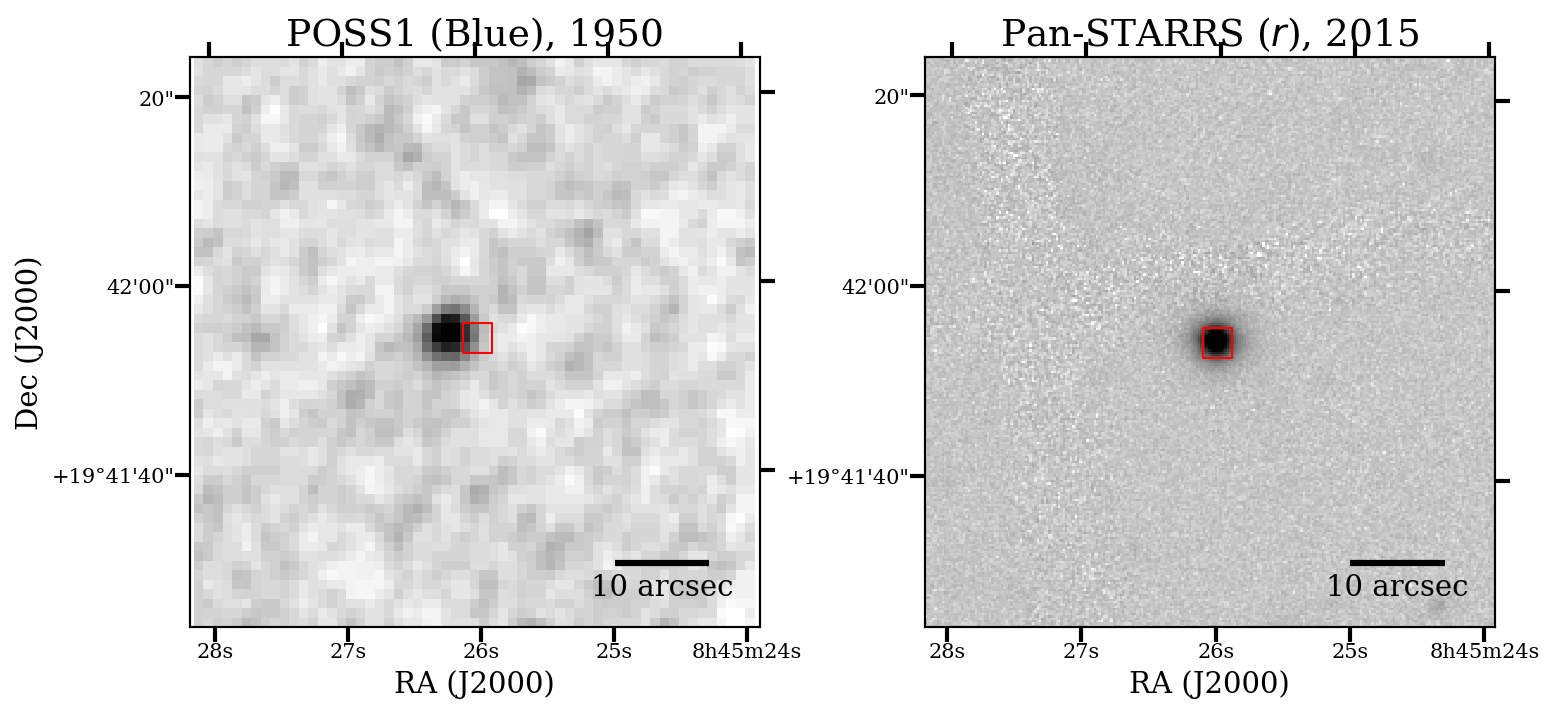}
	\caption{Archival imaging from POSS1 (left) and Pan-STARRS (right), with the position of \target indicated by a red square.}
	\label{fig:poss1}
\end{figure*}

To investigate the possibility of a present-day chance alignment with a background source, we queried 1\arcmin$\times$1\arcmin\, POSS1 images centered on \target from the STScI Digitized Sky Survey.\footnote{\url{http://archive.stsci.edu/cgi-bin/dss_form}} The proper motion of \target is large enough that the imaging from 1950 does not show any hint of a background source at its current position (see \autoref{fig:poss1}).

\subsection{Literature data}

To characterize the host star, we began by gathering literature data, including broadband photometry, astrometry, and physical parameters (see \autoref{tab:stellar}). We sourced the parallax, proper motion, $G$, $B_\mathrm{p}$, $R_\mathrm{p}$ band magnitudes, effective temperature \teff, and radius \rstar of \target from {\it Gaia} DR2 \citep{2016A&A...595A...1G,2018A&A...616A...1G}, as well as optical and infrared photometry from the SDSS \citep{Ahn2012}, Pan-STARRS \citep{Chambers2016}, UKIDSS \citep{Lawrence2007}, 2MASS \citep{2003yCat.2246....0C}, and All{\it WISE} \citep{2013yCat.2328....0C} catalogs.

\section{Analysis}
\label{sec:analysis}

\subsection{Transit modeling}
\label{sec:transit}

To model the transits, we first subtracted long term trends caused by stellar variability or instrument systematics using a cubic spline with knots every 0.75 days. We adopted a Gaussian likelihood function and the analytic transit model of \citet{2002ApJ...580L.171M} as implemented in the Python package {\tt batman} \citep{2015PASP..127.1161K}, assuming a linear ephemeris and quadratic limb darkening. For Markov Chain Monte Carlo (MCMC) exploration of the posterior probability surface, we used the Python package {\tt emcee} \citep{emcee}. To reduce unnecessary computational expense, we only fit the light curves in 4$\times T_{14}$ windows centered on the individual mid-transit times. During MCMC we allowed the free parameters: orbital period $P_\mathrm{orb}$, mid-transit time $T_0$, scaled planet radius $R_p/R_\star$, scaled semi-major axis $a/R_\star$, impact parameter $b\equiv a\cos i/R_\star$, and quadratic limb-darkening coefficients ($q_1$ and $q_2$) under the transformation of \citet{2013MNRAS.435.2152K}. We also fit for the logarithm of the Gaussian errors (log\,$\sigma$) and a constant out-of-transit baseline offset, which was included to minimize any potential biases in parameter estimates arising from the normalization of the light curve. We imposed Gaussian priors on the limb darkening coefficients, with mean and standard deviation determined by Monte Carlo sampling an interpolated grid of the theoretical limb darkening coefficients tabulated by \citet{2012yCat..35460014C}, enabling the propagation of uncertainties in host star effective temperature $T_\mathrm{eff}$, surface gravity log\,$g$, and metallicity [Fe/H] (see \autoref{tab:stellar}).

\begin{figure}
	\includegraphics[clip,trim={0.22cm 0.1cm 0.29cm 0.1cm},width=\columnwidth]{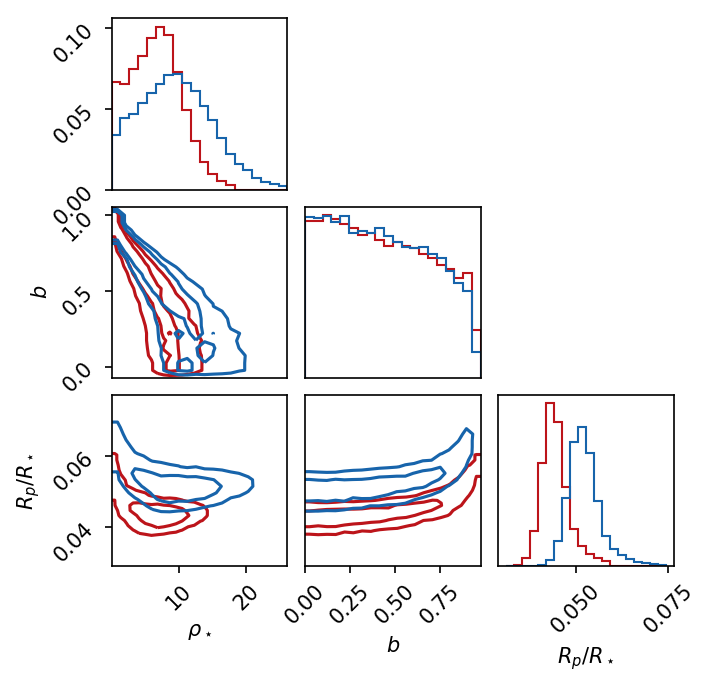}
	\caption{Joint posterior distributions of $\rho_\star$, $b$, and \rp/\rstar without using a prior on stellar density, with 1- and 2-$\sigma$ contours. As in \autoref{fig:k2}, planet b is in red and planet c is in blue.}
	\label{fig:corner}
\end{figure}

We refined initial parameter estimates from BLS by performing a preliminary nonlinear least squares fit using the Python package {\tt lmfit} \citep{newville_2014_11813}, and then initialized 100 ``walkers'' in a Gaussian ball around the least squares solution. We ran MCMC for 5000 steps and visually inspected the chains and posteriors to ensure they were smooth and unimodal, and we computed the autocorrelation time\footnote{\url{https://github.com/dfm/acor}} of each parameter to ensure that we had collected 1000's of effectively independent samples after discarding the first 2000 steps as ``burn-in.'' We also performed transit fits allowing for eccentricity of each planet ($e_\mathrm{b}$ and $e_\mathrm{c}$), but found them to be poorly constrained by the light curve: the upper limits are $e_\mathrm{b} < 0.79$ and $e_\mathrm{c} < 0.87$ (95\% confidence). We show the joint posterior distributions of $\rho_\star$, $b$, and \rp/\rstar for both planets in \autoref{fig:corner}, derived from the MCMC samples obtained as described above. Because we imposed no prior on the mean stellar density, we can confirm that the mean stellar densities derived from the transits of each planet agree with each other and with the density we derive for the host star in \autoref{sec:stellar}. The mean stellar densities from the transit fits of planets b and c are $6.49^{+3.85}_{-4.21}$ and $9.23^{+5.75}_{-5.60}$ \gcc, respectively. These values are in good agreement with each other and with our independent determination of \target's mean stellar density \rhostar = $6.61 \pm 0.32$ \gcc, which provides additional confidence that the observed transit signals both originate from \target. Having confirmed this agreement, we perform a final MCMC analysis assuming a circular orbit and including a Gaussian prior on the mean stellar density. With the exception of the impact parameter $b$, the resulting marginalized posterior distributions appeared symmetric. We report the median and 68\% credible interval of the posteriors in \autoref{tab:planets}; the median and 95\% credible interval of $b$ was $0.40^{+0.26}_{-0.37}$ for planet b, and $0.55^{+0.19}_{-0.47}$ for planet c.

\begin{table}
\scriptsize
\caption{Planet parameters}
\label{tab:planets}
\begin{tabular}{lcll}
\hline
Parameter & Unit & Planet b & Planet c \\
\hline
{\it Free} & & &  \\
$P$ & days & $5.840002^{+0.000676}_{-0.000602}$ & $19.660302^{+0.003496}_{-0.003337}$ \\
$T_{0}$ & BJD & $2458102.59177^{+0.00428}_{-0.00523}$ & $2458117.09169^{+0.00485}_{-0.00447}$ \\
$R_\mathrm{p}$ & \rstar & $0.04318^{+0.00275}_{-0.00259}$ & $0.05164^{+0.00368}_{-0.00354}$ \\
$a$ & \rstar & $22.84^{+0.36}_{-0.38}$ & $51.30^{+0.82}_{-0.84}$ \\
$b$ & --- & $0.40^{+0.16}_{-0.23}$ & $0.55^{+0.12}_{-0.20}$ \\
log($\sigma$) & --- & $-6.89^{+0.06}_{-0.06}$ & $-7.05^{+0.11}_{-0.10}$ \\
$q_1$ & --- & $0.51^{+0.11}_{-0.10}$ & $0.51^{+0.12}_{-0.10}$ \\
$q_2$ & --- & $0.25^{+0.03}_{-0.03}$ & $0.25^{+0.03}_{-0.03}$ \\
\hline
{\it Derived} & & &  \\
\rp & \rearth & $2.231^{+0.151}_{-0.145}$ & $2.668^{+0.201}_{-0.194}$ \\
\teq & K & $496 \pm 10$ & $331 \pm 7$ \\
$a$ & AU & $0.05023^{+0.00042}_{-0.00043}$ & $0.11283^{+0.00095}_{-0.00097}$ \\
$i$ & deg & $89.01^{+0.58}_{-0.40}$ & $89.38^{+0.22}_{-0.13}$ \\
$T_{14}$ & hours & $1.884^{+0.118}_{-0.149}$ & $2.618^{+0.271}_{-0.233}$ \\
$T_{23}$ & hours & $1.701^{+0.137}_{-0.176}$ & $2.256^{+0.325}_{-0.302}$ \\
$R_\mathrm{p,max}$ & \rstar & $0.05114^{+0.01380}_{-0.00825}$ & $0.07426^{+0.02527}_{-0.01822}$ \\
\hline
\end{tabular}
\end{table}

\subsection{Stellar characterization}
\label{sec:stellar}

\begin{figure}
	\includegraphics[clip,trim={0cm 0cm 0cm 0cm},width=\columnwidth]{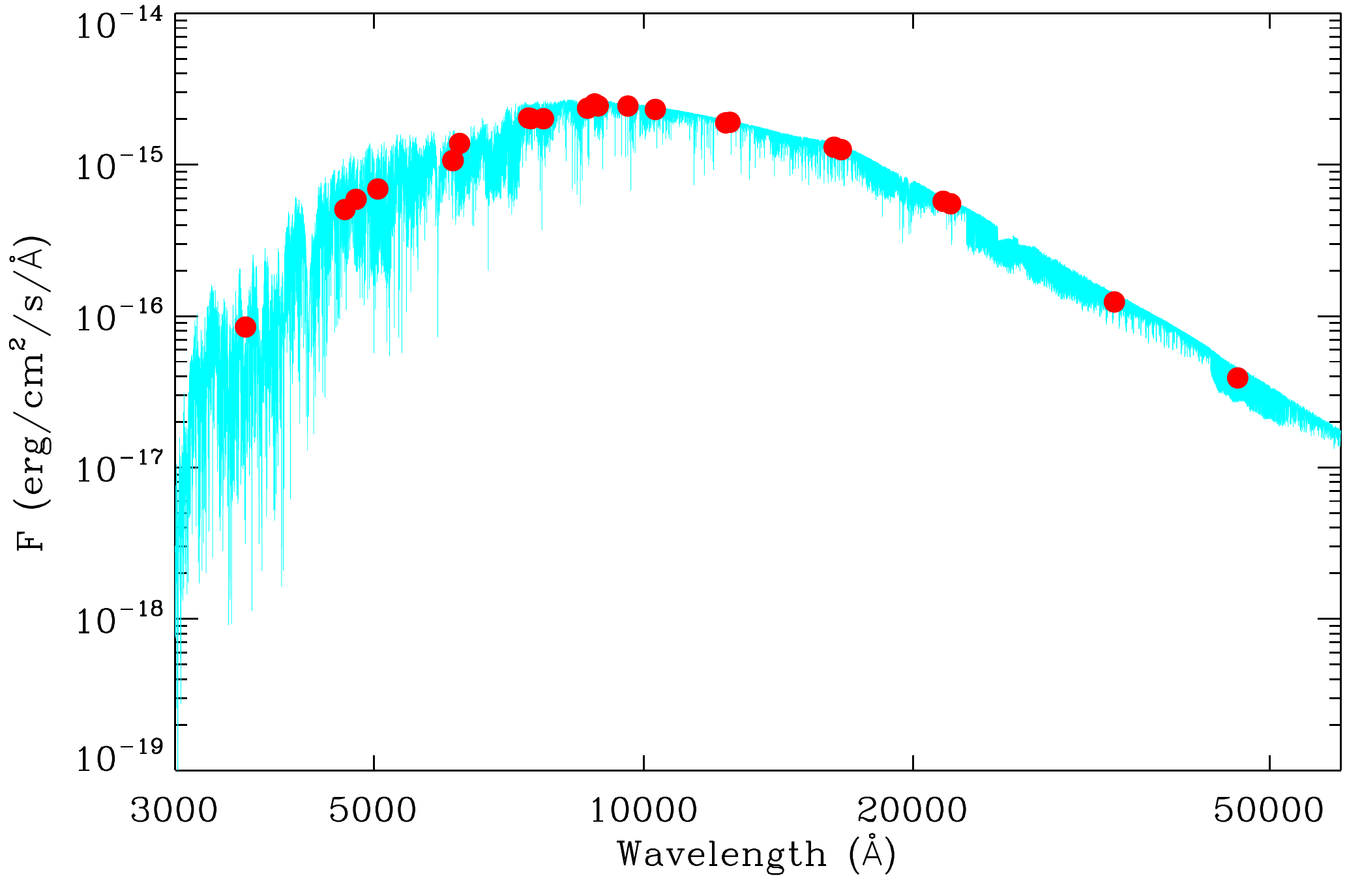}
	\caption{Spectral energy distribution of \target. The red circles mark the observed fluxes as derived from the optical and infrared magnitudes listed in Table\,\ref{tab:stellar}. The best fitting {\tt BT-Settl} is overplotted with a light blue thick line.}
	\label{fig:sed}
\end{figure}

\target is an M2 dwarf star in the Praesepe open cluster \citep{1991AJ....102.1080J, 2007AJ....134.2340K, 2014ApJ...784...57W, 2018A&A...616A..10G}. Estimates for the age of Praesepe lie in the range 600--800 Myr \citep[e.g.][]{2007AJ....134.2340K, 2008A&A...483..891F, 2015ApJ...807...24B}, which is consistent with a recent estimate using data from {\it Gaia} DR2 of log(age) = $8.85^{+0.08}_{-0.06}$ by \citet{2018A&A...616A..10G}. Because the analysis of \citet{2015ApJ...807...24B} accounts for rotation, their older age estimate of $790 \pm 60$ Myr is likely to be more accurate than earlier determinations, but we adopt the full range to be conservative. We note that \target is expected to lie on the main sequence; indeed, stellar evolution models predict an M2 star to reach the main sequence by 150-200 Myr, well before the age of Praesepe.

As a preliminary assessment of the stellar parameters of \target, we built the spectral energy distribution (SED; Fig.\ref{fig:sed}) of \target using the optical and infrared magnitudes listed in Table\,\ref{tab:stellar}. We did not include the All{\it WISE} $W3$ and $W4$ magnitudes because the former has a signal-to-noise ratio of SNR=3.7, while the latter is an upper limit. We used the web-tool {\tt VOSA}\footnote{\url{http://svo2.cab.inta-csic.es/theory/vosa}.} \citep[Version 6;][]{Bayo2008} to compare the SED to the grid of {\tt BT-Settl} synthetic model spectra of very-low-mass stars \citep{Allard2012}. {\tt VOSA} is a virtual observatory tool specifically designed to derive stellar fundamental parameters (e.g., effective temperature, metallicity, gravity, luminosity, interstellar extinction) by comparing the observed SED to theoretical models. We found that \target has an effective temperature of \teff\,=\,$3500\pm 50$\,K, a surface gravity of log\,g\,=\,$5.00 \pm 0.25$ (cgs), and a metallicity of [M/H]\,=\,$0.30\pm0.15$ dex. Assuming a normal value for the total-to-selective extinction ($R=A_\mathrm{v}/E(B-V)=3.1$), we derived an interstellar extinction of $A_\mathrm{v}\,=\,0.03\pm0.03$\,mag. We note that both metal content and extinction are consistent with the average values measured for other member stars of the Praesepe open cluster \citep[see, e.g.,][]{Boesgaard2013,Yang2015}. We used {\it Gaia} DR2 parallax to determine the luminosity and radius of \target. Following \citet{2018A&A...616A...9L}, we accounted for systematic errors in {\it Gaia} astrometry by adding 0.1 mas in quadrature to the parallax uncertainty of \target from {\it Gaia} DR2. Assuming a black body emission at the star's effective temperature, we found a luminosity of $L_\star\,=\,0.0329\pm0.0014$~$L_\odot$ and a radius of $R_\star$\,=\,$0.493\pm0.018$~$R_\odot$.

To obtain the final set of stellar parameters we use in this work, we utilized the \isochrones \citep{2015ascl.soft03010M} Python interface to the {\tt Dartmouth} stellar evolution models \citep{2008ApJS..178...89D} to infer stellar parameters using the 2MASS $JHKs$ photometry and {\it Gaia} DR2 parallax (with augmented uncertainty to account for systematics as above). \isochrones uses the {\tt MultiNest} \citep{2013arXiv1306.2144F} algorithm to sample the posteriors of fundamental stellar properties of interest, and resulted in the following constraints: effective temperature \teff = $3660^{+80}_{-45}$ K, surface gravity \logg\,=\,$4.783 \pm 0.012$ (cgs), metallicity \feh\,=\,$-0.013 \pm 0.180$ dex, radius \rstar\,=\,$0.473 \pm 0.011$ \rsun, mass \mstar\,=\,$0.496 \pm 0.013$ \msun, extinction ($A_\mathrm{V}$)\,=\,$0.301 \pm 0.162$ mag, and distance\,=\,$187.0 \pm 4.0$ pm. We opted not to include a prior on the metallicity of \target based on its cluster membership, as the resulting stellar parameter uncertainties may not accurately reflect intrinsic variability of metallicity within the Praesepe birth nebula. The posteriors agree with the results of our SED analysis to within $\sim$2$\sigma$ and are consistent with Praesepe membership; metallicity is poorly constrained, but is consistent with that of Praesepe \citet[\feh = $0.12 \pm 0.04$;][]{Boesgaard2013}. This mild disagreement is likely the result of systematics from the underlying stellar models, which are unaccounted for in the formal uncertainties. Most posteriors appeared roughly symmetric and Gaussian, so we list the median and standard deviation in \autoref{tab:stellar}; the \teff posterior was asymmetric, so we list the median and 68\% credible region instead. We note that these values are in moderate disagreement with the stellar parameters computed by \citet{2016ApJS..224....2H}, which may be due to the lack of a parallax constraint in their analysis, but may also reflect a systematic bias for low mass stars, which has been attributed to their choice of stellar models \citep{2017ApJ...836..167D}. We note that these estimates are consistent with the {\it Gaia} DR2 values for \target (\teff\,=\,3422$^{+478}_{-22}$ K, \rstar\,=\,0.54$^{+0.01}_{-0.12}$ \rsun, distance\,=\,$186.573 \pm 2.105$ pc).

\begin{figure}
	\includegraphics[clip,trim={0cm 0cm 0cm 0cm},width=0.9\columnwidth]{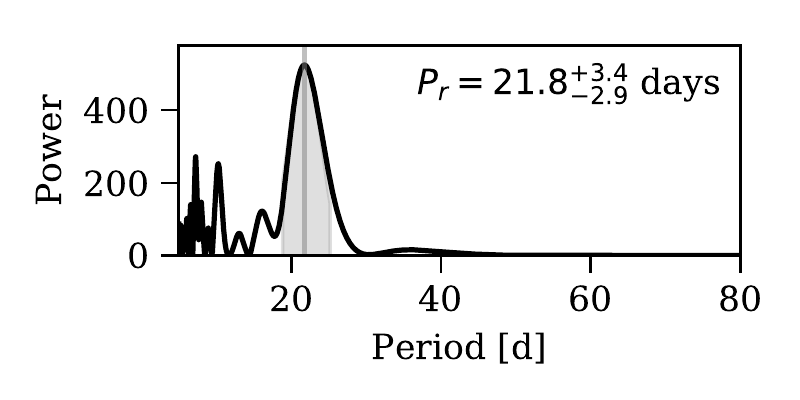}
	\caption{Lomb-Scargle periodogram of the K2 light curve of \target.}
	\label{fig:ls-rot}
\end{figure}

\begin{figure*}
	\includegraphics[clip,trim={0cm 0cm 0cm 0cm},width=0.99\textwidth]{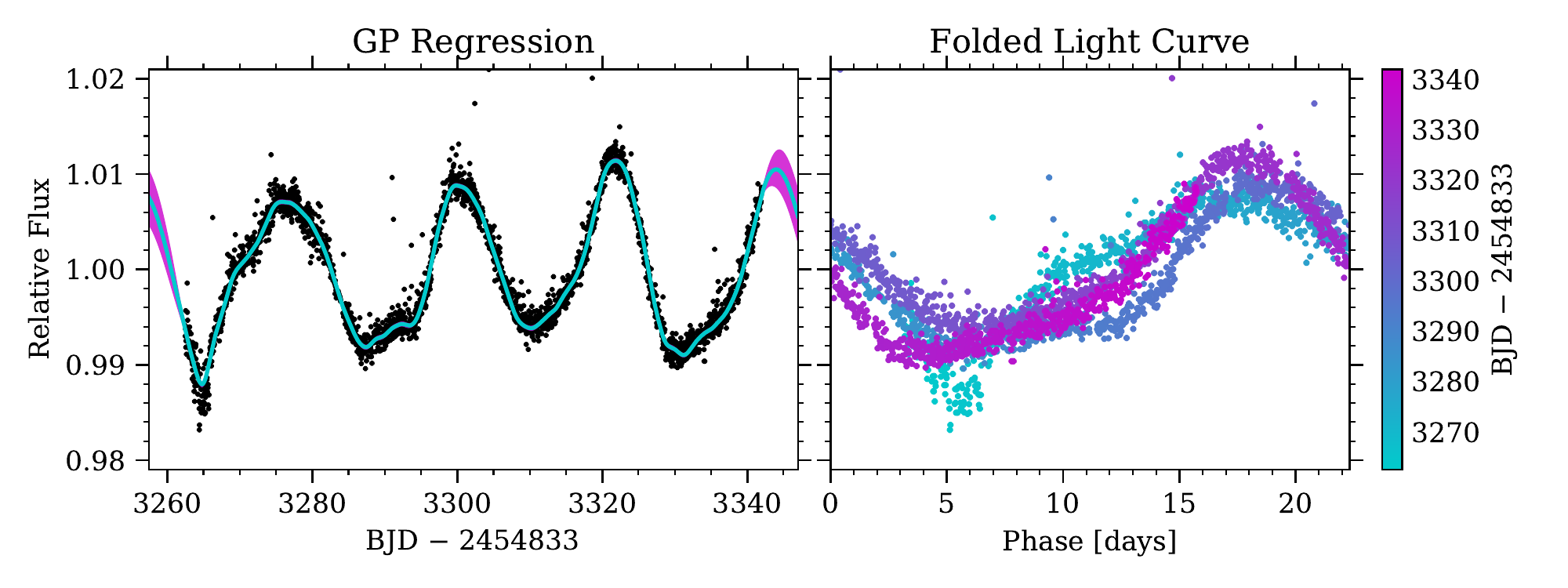}
	\caption{Gaussian Process fit to the light curve of \target with transits removed (left), and the same light curve folded on the maximum a posteriori rotation period of 22.3 days (right; color of datapoints correspond to time).}
	\label{fig:gp-rot}
\end{figure*}

The light curve of \target exhibits clear quasi-periodic rotational modulation, which is characteristic of surface magnetic activity regions moving in and out of view as the star rotates around its axis. We measured the rotation period using two different methods. After masking the transits from the \ktwo light curve and subtracting a linear trend, we computed the Lomb-Scargle periodogram, from which we derived a stellar rotation period of $21.8^{+3.4}_{-2.9}$ days by fitting a Gaussian to the peak (see \autoref{fig:ls-rot}). In \autoref{fig:gp-rot} we show a Gaussian Process (GP) fit to the light curve using a quasi-periodic kernel \citep[e.g.][]{2014MNRAS.443.2517H, 2015ApJ...808..127G, 2017AJ....154..226D}, from which we measured a rotation period of $22.2 \pm 0.6$ days via MCMC exploration of the kernel hyperparameter space. We adopt the GP estimate, as it is in good agreement with the Lomb-Scargle estimate but yields higher precision. The rotational modulation of \target has a similar period and amplitude to K2-95, an M3 dwarf in Praesepe hosting a transiting sub-Neptune \citep{2016AJ....152..223O}. Using the gyrochronology relation of \citet{2015MNRAS.450.1787A} we find that the measured stellar rotation period is consistent with the age of Praesepe.

\begin{table}
\caption{Stellar parameters}
\label{tab:stellar}
\begin{tabular}{lccr}
\hline
Parameter & Unit & Value & Source \\
\hline
{\it Astrometry} &  &  & \\
$\alpha$ R.A.			& deg				& 131.358352378				& {\it Gaia} DR2 \\
$\delta$ Dec.			& deg				& 19.698400987		& {\it Gaia} DR2 \\
$\pi$			& mas 				& $5.3598 \pm 0.0605$		& {\it Gaia} DR2 \\
$\mu_\alpha$	& mas\,yr$^{-1}$ 	& $-37.900 \pm 0.095$		& {\it Gaia} DR2 \\
$\mu_\delta$	& mas\,yr$^{-1}$ 	& $-13.079 \pm 0.061$		& {\it Gaia} DR2 \\
\hline
{\it Photometry} &  &  & \\
$Kp$			& mag 				& 15.318			 		& EPIC \\
\noalign{\smallskip}
$B_\mathrm{p}$            & mag               & $16.946 \pm 0.006$        & {\it Gaia} DR2 \\
$R_\mathrm{p}$            & mag               & $14.538 \pm 0.002$        & {\it Gaia} DR2 \\
$G$				& mag 				& $15.663 \pm 0.001$		& {\it Gaia} DR2 \\
\noalign{\smallskip}
$u$             & mag               & $19.994 \pm 0.036$          & Sloan/SDSS \\
$g$             & mag               & $17.499 \pm 0.005$          & Sloan/SDSS \\
$r$             & mag               & $16.089 \pm 0.004$          & Sloan/SDSS \\
$i$             & mag               & $14.963 \pm 0.004$          & Sloan/SDSS \\
$z$             & mag               & $14.374 \pm 0.004$          & Sloan/SDSS \\
\noalign{\smallskip}
$g$             & mag               & $17.260 \pm 0.006$          & Pan-STARRS \\
$r$             & mag               & $16.075 \pm 0.002$          & Pan-STARRS \\
$i$             & mag               & $14.965 \pm 0.003$          & Pan-STARRS \\
$z$             & mag               & $14.471 \pm 0.002$          & Pan-STARRS \\
$y$             & mag               & $14.221 \pm 0.004$          & Pan-STARRS \\
\noalign{\smallskip}
$Z$             & mag               & $13.848 \pm 0.002$          & UKIDSS     \\
$J$             & mag               & $12.997 \pm 0.002$          & UKIDSS     \\
$H$             & mag               & $12.393 \pm 0.001$          & UKIDSS     \\
$K$             & mag               & $12.157 \pm 0.001$          & UKIDSS     \\
\noalign{\smallskip}
$J$ 			& mag 				& $13.047 \pm 0.025$ 		& 2MASS \\
$H$				& mag			 	& $12.386 \pm 0.022$		& 2MASS \\
$Ks$			& mag 				& $12.183 \pm 0.020$			& 2MASS \\
\noalign{\smallskip}
$W1$			& mag 				& $12.048 \pm 0.023$			& All{\it WISE} \\
$W2$			& mag 				& $11.978 \pm 0.023$			& All{\it WISE} \\
$W3$            & mag               & $11.317 \pm 0.294$            & All{\it WISE} \\
$W4$            & mag               & $8.173$                       & All{\it WISE} \\
\hline
{\it Physical} &  &  & \\
\teff 			& K 				& 3660$^{+80}_{-45}$ 		& This work \\
\logg 			& cgs 				& $4.783 \pm 0.012$ 	    & This work \\
\feh  			& dex 				& $-0.013 \pm 0.180$ 	    & This work \\
\mstar 			& \msun 			& $0.496 \pm 0.013$ 	    & This work \\
\rstar 			& \rsun	 			& $0.473 \pm 0.011$         & This work \\
\rhostar        & \gcc              & $6.610 \pm 0.322$         & This work \\
$A_\mathrm{V}$  & mag 				& $0.301 \pm 0.162$       	& This work \\
distance        & pc 				& $187.0 \pm 4.0$ 	        & This work \\
$P_\mathrm{r}$  & days              & $22.2 \pm 0.6$			& This work \\
\hline
\end{tabular}
\end{table}

\subsection{Validation}

Transiting planet false positive scenarios typically involve an eclipsing binary (EB) blended with a brighter star within the photometric aperture. If an EB's mass ratio is close to unity, then the primary and secondary eclipses will have the same depth, and in such a case the dilution from the brighter star will make these eclipses shallower and thus more similar to planetary transits. In such a scenario, the EB's orbit must also be circular such that the eclipses mimic the regular periodicity of planetary transits. Another possibility is an extreme EB mass ratio, in which case the (diluted) secondary eclipses would be small enough that they could be below the detection limit of the photometry. Because of the large (4\arcsec) pixel scale of the \kepler photometer, blended EB scenarios are not rare, and must therefore be properly accounted for. Such a false positive scenario could be caused by the chance alignment of a background source (BEB), or by a hierarchical triple system (HEB), the relative frequencies of which depend on the density of sources in the vicinity of the candidate host star.

To investigate the possibility of a BEB false positive scenario, we utilize the observed transit geometry in conjunction with a simulated stellar population appropriate for the line of sight to \target. The eclipse depth of an EB can in principle reach a maximum of 100\%, which sets a limit on the faintness of any putative background sources that could be responsible for the observed signals. Using Equation 1 of \citet{2018AJ....156...78L} and the observed transit depths, this corresponds to $Kp \approx 22$ mag. Using a simulated stellar population in the direction of \target from {\tt TRILEGAL} Galaxy model \citep{2005A&A...436..895G}, the expected frequency of sources brighter than this limit is very low, at $\sim$0.07 for a 7\arcsec\, photometric aperture (see \autoref{fig:aper}). Indeed, the non-detection of any background sources in our AO image (see \autoref{fig:ao}) and the POSS1 image from 1950 (see \autoref{fig:poss1}) is consistent with the expectation of zero such sources from the Galaxy model.

If, on the other hand, the observed signals are actually the result of a HEB scenario, we must instead consider the possibility that \target is actually a bound triple star system. In order for the eclipsing component to have a negligible impact on the observed SED (see \autoref{fig:sed}), it would need to be composed of stars with much lower masses than \target. However, from the observed transit geometry we have 3-$\sigma$ upper limits on the radius ratio of 9\% and 15\% for the inner and outer planets, respectively, using Equation 21 of \citet{2003ApJ...585.1038S} (see $R_\mathrm{p,max}$ in \autoref{tab:planets}). Radius ratios below this limit would involve either an eclipsing component in the planetary mass regime or an occulted component that would contribute non-negligible flux to the combined SED and thereby have observable signatures. Perhaps most importantly, the existence of two periodic transit-like signals from the same star is {\it a priori} more difficult to explain with non-planetary scenarios, because the BEB and HEB scenarios consistent with the observed signals would require vanishingly infrequent chance alignment or higher stellar multiplicity. Indeed, candidates in systems of multiple transiting planets have been shown to have a very low false positive rate \citep{2012ApJ...750..112L}, and are thus essentially self-validating.

Besides these qualitative considerations, we also computed the false positive probabilities (FPPs) of the planet candidates of \target using the Python package \vespa \citep{2015ascl.soft03011M}. \vespa employs a robust statistical framework to compare the likelihood of the planetary scenario to likelihoods of several astrophysical false positive scenarios involving eclipsing binaries, relying on simulated eclipsing populations based on {\tt TRILEGAL}. The FPPs from \vespa for planets b and c are 0.007\% and 0.012\%, respectively, well below the standard validation threshold of 1\%. Moreover, these FPPs are overestimated due to the fact that \vespa does not account for multiplicity: \citet{2012ApJ...750..112L} demonstrated that a candidate in a system with one or more additional transiting planet candidates is 25 times more likely to be a planet based on multiplicity alone. Therefore, in addition to the qualitative arguments above, the planet candidates also quantitatively warrant validation; we conclude that \target is thus the host of two {\it bona fide} transiting planets.

\subsection{Dynamical stability}

Given the large separation between the two planets, the system is manifestly Hill stable. Assuming that the orbits are circular, their separation is about $25$ times their mutual Hill radius, much larger than the threshold value of $3.46 R_\mathrm{H}$ \citep{1993Icar..106..247G,1996Icar..119..261C,2013ApJ...774..129D}. Using the angular momentum deficit criterion of \citet{2018arXiv180608869P}, we find that the eccentricity of the outer planet must be less than $e_\mathrm{c} \simeq 0.4$ to ensure the stability of the system.

\begin{figure}
	\includegraphics[width=0.47\textwidth]{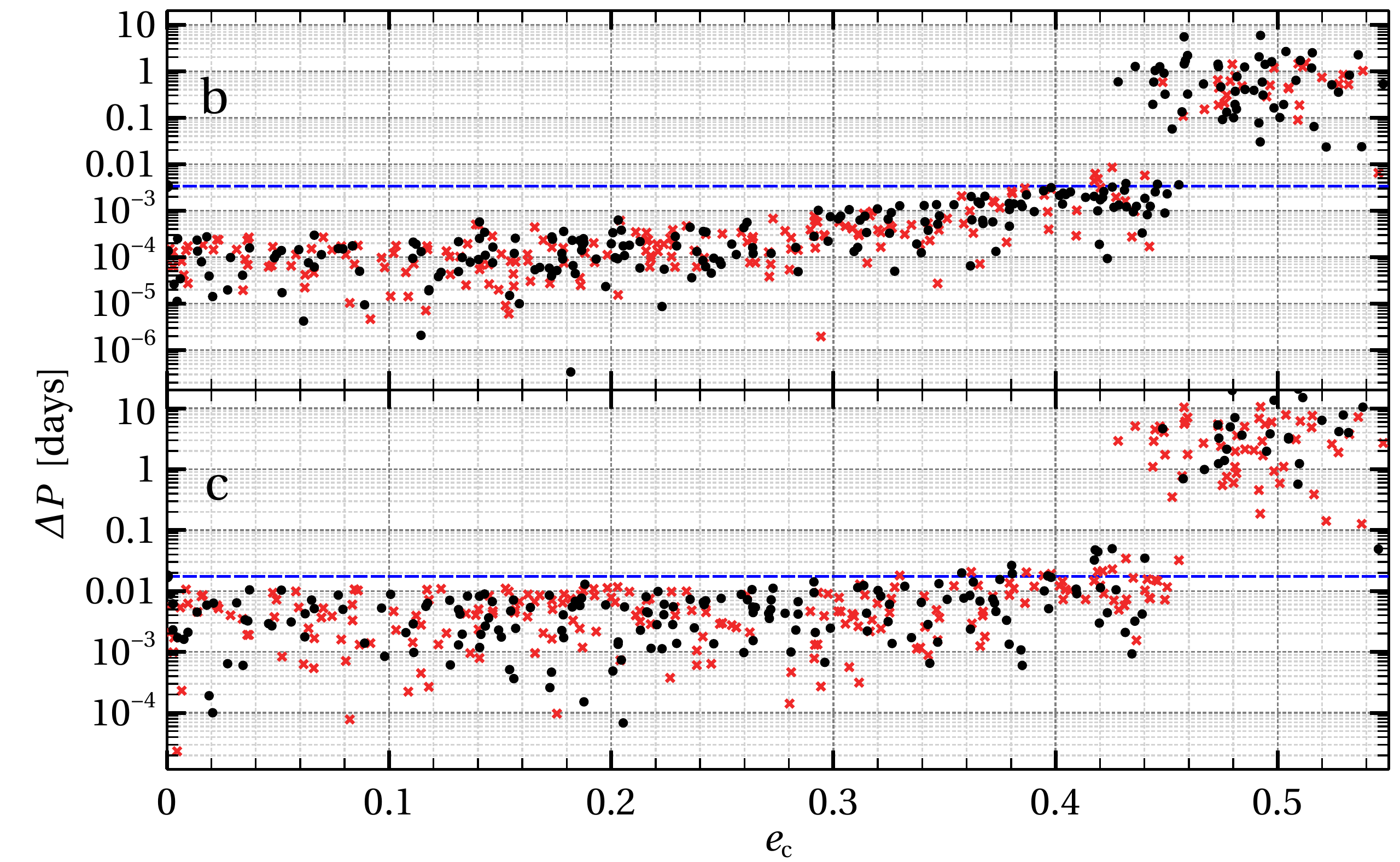}
	\caption{Difference between the initial and final orbital periods of the planets in the $N$-body simulations as a function of the initial eccentricity of the outer planet c. Top panel: period difference of the inner planet b. Bottom panel: period difference of the outer planet c. Black circles and red crosses represent a period increase and decrease with respect to the initial one, respectively. The dashed blue line is the $5\sigma$ error on the period from \autoref{tab:planets}.}
	\label{fig:stab}
\end{figure}

We use the probabilistic mass-radius relation of \citet{2016ApJ...825...19W} to estimate the masses of the planets given their measured radii, yielding $m_\textrm{b} = 7.7 \pm 2.3$ and $m_\textrm{c} = 9.5 \pm 2.7$ \mearth for planets b and c, respectively. We use the {\tt TSUNAMI} code \citep{tra2016ApJ...831...61T,tra2018arXiv180907339T}
to simulate the orbital evolution of 500 realizations. Consistent with the planets' orbital inclinations from the measured transit geometry, we set their mutual inclination to zero and sampled the eccentricity of the outer planet between $0$ and $0.6$. \autoref{fig:stab} shows the difference between the initial orbital periods and the final ones, after $2\,\textrm{Myr}$ of integration. For all systems with $e_\textrm{c} \lesssim 0.3$ the difference in orbital periods remain below $0.01$ days. On the other hand, for $e_\textrm{c} \gtrsim 0.45$, the perturbations between the two planets lead to instability and the period changes significantly ($\Delta P \approx 1$ day). Therefore, $e_\textrm{c} \simeq 0.43$ is a robust upper limit for the eccentricity of the outer planet.

\begin{figure}
	\includegraphics[width=0.47\textwidth]{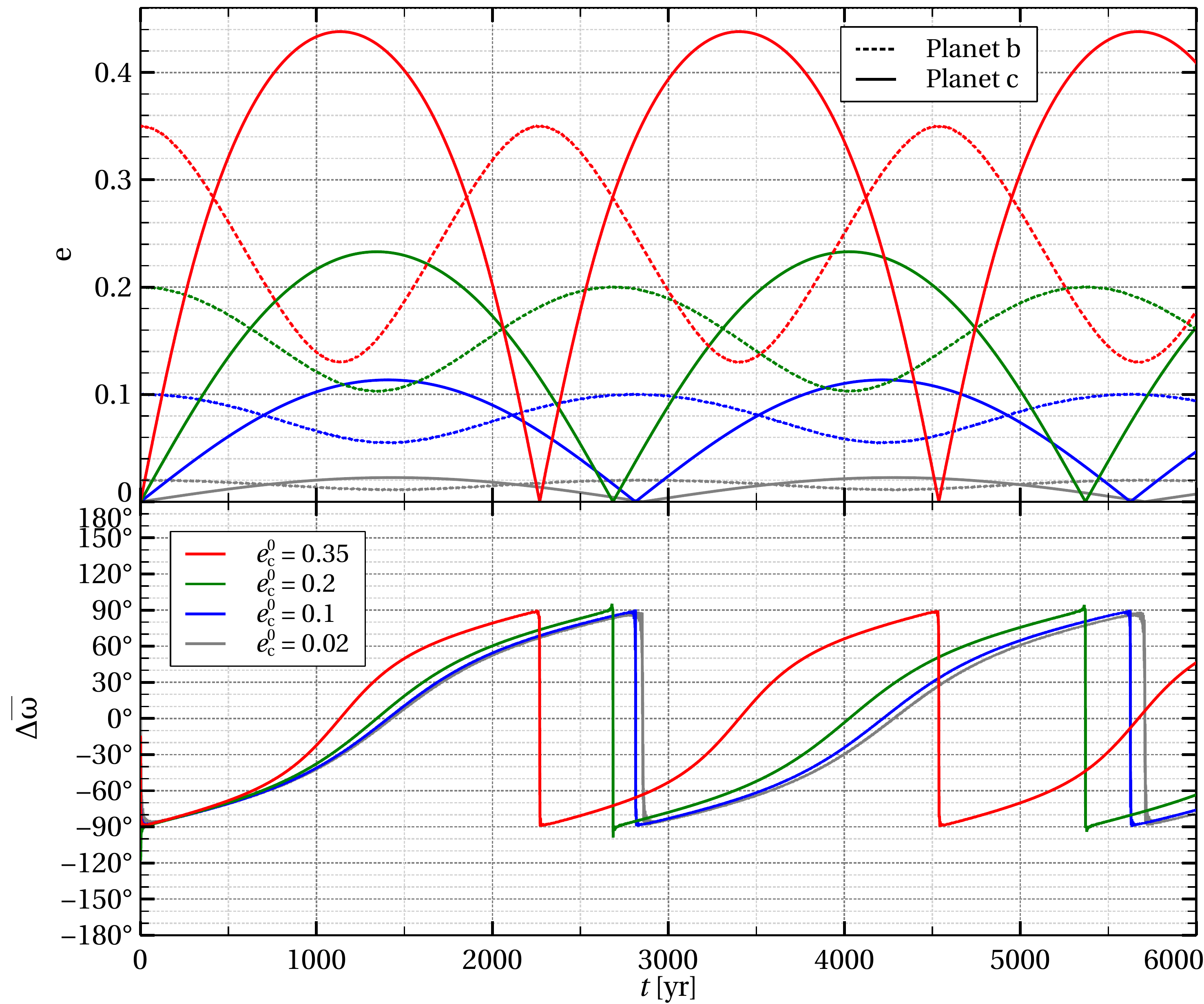}
	\caption{Eccentricity of the planets (top) and difference of longitude of pericenter $\Delta\bar{\omega}$ (bottom) as a function of time, for 4 realization with different initial eccentricity of planet c. Each color is a different realization: red lines $e^0_\textrm{c} = 0.35$; green lines $e^0_\textrm{c} = 0.2$; blue lines $e^0_\textrm{c} = 0.1$; gray lines $e^0_\textrm{c} = 0.02$. In the top panel, the dashed and solid lines are the eccentricity of planet b and c, respectively.}
	\label{fig:osc}
\end{figure}

However, we find that for any eccentricity of the outer planet, the planets undergo secular exchanges of angular momentum which cause the eccentricity of the planets to oscillate periodically (top panel of \autoref{fig:osc}). The outer planet oscillates between the initial eccentricity and a lower value, while the inner one oscillates between 0 and an upper value $e_\mathrm{b}^\mathrm{max}$, which depends on $e_\textrm{c}^0$. We find that $e^\mathrm{max}_\textrm{b}$ and $e_\textrm{c}^0$ are nicely fit, with almost no scatter, by the superlinear relation $e^\mathrm{max}_\textrm{b} = e_\textrm{c}\,(1.12 \,+\, 0.42\,e_\textrm{c})$. Each oscillation has a period of about $2250$--$2800 \rm \, yr$, depending on $e_\textrm{c}^0$. We have also run some tests using different masses of the planets, in the $1\sigma$ error range derived by the mass-radius relation. The eccentricity oscillations occur also for different planet masses, with the oscillation period becoming longer for decreasing mass ratio $m_\textrm{c}/m_\textrm{b}$.

This secular behavior has been found also in other eccentric multiplanet systems \citep[e.g.][]{kane14,barnes06b}. In particular, our system lies on near a boundary between
libration and circulation \citep{barnes06a}. For any initial eccentricity of the planet c, the angle between the two apsidal lines ($\Delta \bar{\omega}$) shows libration between $90^\circ$ and $-90^\circ$ and a rapid change when the inner planet becomes circular (bottom panel of \autoref{fig:osc})

Therefore, we calculate the eccentricity damping timescale for planet b using the tidal model of \citet{1981A&A....99..126H}. Since \target is a low mass star with a convective envelope, we can compute the the tidal timescale $k/T$ due to the tides raised on the star from the stellar structure parameters \citep{1977A&A....57..383Z}.
We use the stellar models of the PARSEC stellar evolution code \citep{bressan2012,chen2015,2018MNRAS.476..496F}, and derive $k/T = 0.34\,\rm yr^{-1}$. Considering only the tides raised on the star, we find that the circularization timescale is much longer than the age of the system for all $e_\textrm{b}<0.9$. We also take into account the tides raised on the planet c using the tidal quality factor $Q = n\,T/3\,k$, where $n$ is the mean motion of planet c. Note that the stellar $k/T$ corresponds to a quality factor of ${\sim}350$, which tells us that any $Q > 350$ makes the planetary tide less efficient the stellar tide. Since the tidal quality factor for gas giants is expected to be $Q\gg 10^2$, we conclude that the inner planet could be still undergoing tidal circularization.

\section{Discussion}
\label{sec:discuss}

\begin{figure}
	\includegraphics[clip,trim={0.35cm 0cm 0.4cm 0.25cm},width=1\columnwidth]{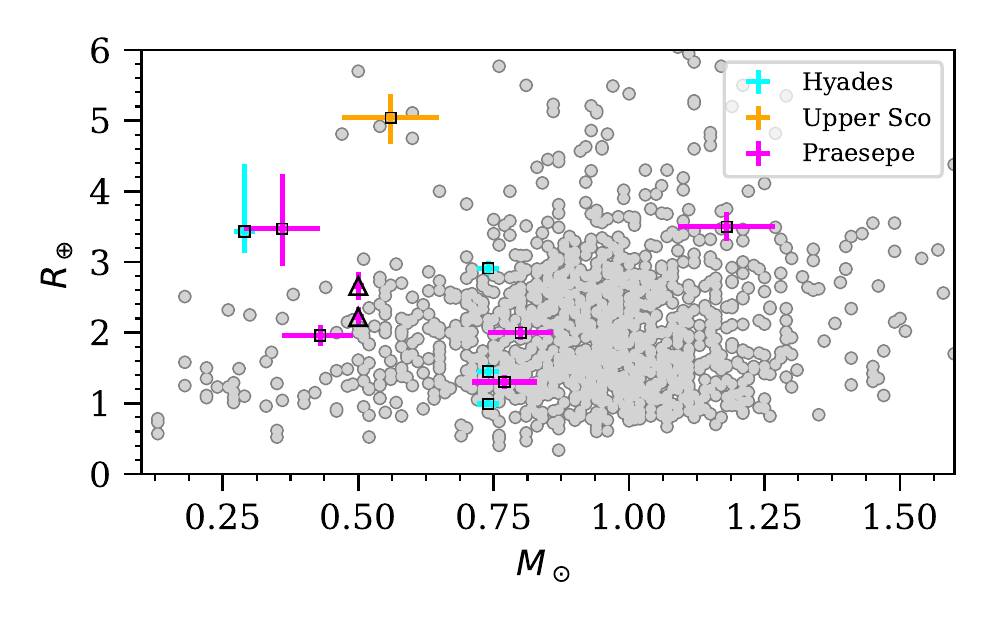}
	\caption{Planet radius versus host star mass of \target (triangles) and a selection of other transiting planet systems in clusters (squares), as compared to the field star planet population (gray points). Besides \target, the data shown are from a query of the NASA Exoplanet Archive \citep{2013PASP..125..989A}. Besides \target, the cluster systems shown are: K2-25 and K2-136 \citep[Hyades;][]{2016ApJ...818...46M, 2018AJ....155....4M, 2018AJ....155..115L, 2018AJ....155...10C}; K2-33 \citep[Upper Sco;][]{2016Natur.534..658D, 2016ApJ...818...46M}; K2-95, K2-100, K2-101, K2-102, K2-103, and K2-104 \citep[Praesepe;][]{2016AJ....152..223O, 2017AJ....153...64M} }
	\label{fig:rp-ms}
\end{figure}

Assuming a bond albedo of 0.3, the equilibrium temperatures of planets b and c are $496 \pm 13$ and $331 \pm 8$ K, respectively, making \target a system of two warm sub-Neptunes. Although such planets have been found in large numbers by previous surveys (e.g. \kepler), the number orbiting cluster stars is extremely small. \target is thus an important system because it significantly improves the statistics for demographic studies of cluster planets. Furthermore, prior to this discovery only one member of a cluster was known to host multiple transiting planets \citep[K2-136;][]{2018AJ....155....4M,2018AJ....155..115L,2018AJ....155...10C}. \target is thus a unique laboratory for studies of system architectures as a function of time. We place \target in the context of the general exoplanet population, as well as other cluster planets, by plotting planet radius as a function of host star mass in \autoref{fig:rp-ms}, using data from a query of the NASA Exoplanet Archive \footnote{\url{https://exoplanetarchive.ipac.caltech.edu/}} \citep{2013PASP..125..989A}. From this perspective, the planets of \target do not appear to have significantly inflated radii, as has previously been a matter of speculation for cluster systems (e.g. K2-25, \citealt{2016ApJ...818...46M};  K2-95, \citealt{2016AJ....152..223O}). It is worth noting, however, that K2-25 and K2-95 have lower masses than \target, and the radii of planets orbiting higher mass host stars in both Hyades and Praesepe appear less inflated. Two cluster planets buck this apparent trend: K2-33 and K2-100. However, K2-33 may still be undergoing radial Kelvin–Helmholtz contraction due to its young age \citep[5--10 Myr;][]{2016Natur.534..658D}, and K2-100 is much more massive \citep[\mstar = $1.18 \pm 0.09$ \msun;][]{2017AJ....153...64M}. The radii of the planets orbiting \target lend support to this trend, and thus to the hypothesis that radius inflation results from higher levels of X-ray and ultraviolet (UV) flux incident upon planets orbiting lower mass stars; the absence of such a trend for field stars may tell us something about the timescales of radial relaxation after early-stage X-ray/UV flux from low mass stars diminishes.

Planets orbiting cluster stars are expected to have large eccentricities and large mutual inclinations, if perturbations from cluster members are efficient. While we cannot yet constrain the eccentricity of the outer planet, it is safe to assume that the system is coplanar. Even in the hypothesis of the presence of an outer, inclined, non-transiting planet, produced during a stellar encounter in the early life of the cluster, perturbations from the outer planet would have propagated inward, altering the the inclinations of the inner planets. \citet{2018MNRAS.474.5114C} show that planetary systems in the outskirts of the cluster (i.e. outside its half-mass radius) are unlikely to have been perturbed by passing stars. We compute the distance of \target from the center of Praesepe, using the cluster center coordinates derived by \citet{2013MNRAS.434.3236K} and the coordinates of \target from {\it Gaia} DR2. We find that \target lies at $4.365\pm 0.206\,\textrm{pc}$ projected distance ($8.8 \pm 4.2$ unprojected) from the cluster center, well outside the half-mass radius of the Praesepe cluster ($3.9\,\textrm{pc}$, \citealt{2013MNRAS.434.3236K}). This suggests that perturbations from other stars have likely played a minor role in shaping the planetary system of \target.

Because the planets orbiting \target have a common host star history, X-ray and UV stellar flux at young ages can be controlled for, better enabling their observed radii to yield insights into atmospheric evolution due to irradiation from the host star. Additionally, the 600--800 Myr age of the system is particularly good for testing photoevaporation theory, as this is the timescale over which photoevaporation should have finished \citep{2013ApJ...775..105O}; by this age, the radius distribution of small planets should approach that of field stars. The planet radii place them both securely above the observed gap in the radius distribution \citep{2017AJ....154..109F, 2018MNRAS.479.4786V, 2018arXiv180500231B}, which suggests either that they have large enough core masses to have retained substantial atmospheres, or that photoevaporation may have played a less significant role in their evolution, or both. However, the host star's spectral type indicates substantial X-ray/UV irradiation during the first few hundred million years, which makes it less likely that the planets could have largely escaped the effects of photoevaporation unless they had larger core masses. Indeed, the location of the bimodality has been shown to shift to smaller radii for lower mass host stars \citep{2018arXiv180501453F}, consistent with the expectation that smaller stars produce smaller planet cores. This implies that the planets orbiting \target are more likely to have relatively massive cores and always occupied the larger radius mode. Given the age of the system, it is likely that photoevaporation is effectively over, and the planet radii will no longer undergo substantial evolution.

Systems of multiple transiting planets sometimes allow for the masses and eccentricities in the system to be measured via dynamical modeling of the observed transit timing variations \citep[TTVs;][]{2005Sci...307.1288H, 2005MNRAS.359..567A}, in which the mutual gravitational interaction between planets produces regular, measurable deviations from a linear ephemeris. To test if either planet exhibits TTVs, we used the best-fitting transit model as a template for the determination of individual transit times. Keeping all parameters fixed except the mid-transit time, we fitted this template to each transit in the data, but we did not detect any TTVs over the $\sim$80 days of \ktwo observations. The absence of TTVs is perhaps not surprising given that the orbital periods are not especially close to a low-order mean motion resonance, with $P_c / P_b \approx 3.367$, about 12\% outside of a 3:1 period commensurability. Although the planets do not exhibit measurable mutual gravitational interactions, the pull they exert on their host star presents an opportunity for characterization via Doppler spectroscopy.

By obtaining precise RV measurements of \target, it may be possible to measure the reflex motion of the host star induced by the gravity of its planets \citep[e.g.][]{1952Obs....72..199S, 1995Natur.378..355M}. Such measurements would yield the planet masses and mean densities, which would constrain the planets' interior structures. The predicted masses of planets b and c, along with their orbital periods and the mass of the host star, yield expected RV semi-amplitudes values of $4.4 \pm 1.3$ and $3.6 \pm 1.0$ m\,s$^{-1}$, respectively. However, the youth and photometric variability of \target imply RV stellar activity signals larger in amplitude than the expected planet signals from optical spectroscopy. This suggests that the planets of \target may be amenable to mass measurement using a high precision NIR spectrograph, such as IRD \citep{2012SPIE.8446E..1TT} or HPF \citep{2012SPIE.8446E..1SM}, as the RV amplitude of stellar activity signals should be significantly lower in the infrared. Assuming no orbital obliquity, from the radius and rotation period of \target we estimate low levels of rotational line broadening, with \vsini of $\sim$1 km\,s$^{-1}$. Prior knowledge about the star's rotation period from \ktwo should prove useful for modeling the stellar activity signal simultaneously with the Keplerian signals of the planets using a Gaussian Process model \citep[e.g.][]{2014MNRAS.443.2517H, 2015ApJ...808..127G, 2017AJ....154..226D}.

Besides spectroscopy, follow-up NIR transit photometry of \target could enable a better characterization of the system by more precisely measuring the transit geometry. Besides yielding a better constraint on the planet radius, transit follow-up would also significantly refine estimates of the planets' orbital ephemerides, enabling efficient scheduling for any subsequent transit observations, e.g. with \jwst. Using the \wise W2 magnitude in \autoref{tab:stellar} as a proxy for \spitzer IRAC2, the expected transit SNR is in the range 4--8; given the systematic noise in \spitzer light curves, such transit measurements would be challenging, but may be feasible by simultaneously modeling the transit and systematics signals using methods such as pixel-level decorrelation \citep[PLD;][]{2015ApJ...805..132D}. Furthermore, by simultaneously modeling the \ktwo and \spitzer data, \spitzer's high photometric observing cadence and the diminished effects of limb-darkening in the NIR could be leveraged to more precisely determine the transit geometry \citep{Livingston_2019}. NIR transit observations from the ground could also be useful, but would likely require a large aperture (e.g. 4--8 meter) telescope to yield better performance than \spitzer.

This system was also reported by \citet{2018arXiv180807068R}, who performed an independent analysis using a \ktwo light curve produced by a different pipeline \citep[K2SFF;][]{2014PASP..126..948V}, as well as follow-up medium-resolution NIR spectroscopy. The estimates of \teff, \feh, \rstar, and \rhostar all agree to within 1$\sigma$, whereas the \mstar estimates differ by 1.4$\sigma$; this mild tension in mass likely reflects the underlying model dependency in our \isochrones analysis. However, \rstar is in perfect agreement between the two analyses; robustness in this parameter is crucially important for measurement of the planet radii. Finally, the estimates of orbital period and \rp/\rstar agree within 1$\sigma$; we thus find 1$\sigma$ agreement for the planet properties \rp and \teq. Our analysis of the publicly available K2SFF light curve\footnote{\url{https://archive.stsci.edu/prepds/k2sff/}} also yields parameters in good agreement. However, the systematics in both light curves were corrected using very similar techniques, so we performed an additional check to see if residual red noise could be significantly affecting our parameter estimates. To do so, we used a GP with a Mat\'ern-3/2 kernel to model the covariance structure of the noise in conjunction with the transits; we found planetary parameters within 1-sigma of the values found previously, suggesting low levels of residual red noise. Taken together, these two independent studies reinforce one another, suggesting a high degree of reliability in the properties of the system.

\section{Summary}
\label{sec:summary}

Using data from the \ktwo mission and ground-based follow-up observations, we have detected and statistically validated two warm sub-Neptunes transiting the star \target, which is a member of the 600-800 Myr Praesepe open cluster. Unlike several previously discovered planets orbiting lower mass stars in clusters, their radii are fairly consistent with the those of planets orbiting field stars of comparable mass to their host, suggesting that radius inflation is a function of host star mass. The system presents opportunities for RV follow-up using high precision NIR spectrographs, which would yield the planets' densities and thereby test theories of planet formation and evolution. NIR transit photometry could more precisely measure the planet's ephemerides and transit geometry, and thus also their radii. By leveraging the known age of the system, such characterization would yield a direct view of the planets' atmospheric evolution. \target joins a small but growing list of cluster planets, and is particularly valuable as it is only the second known system of multiple transiting planets in a cluster.

\section*{Acknowledgements}

This work was carried out as part of the KESPRINT consortium. J.\,H.\,L. gratefully acknowledges the support of the Japan Society for the Promotion of Science (JSPS) Research Fellowship for Young Scientists. This work was supported by Japan Society for Promotion of Science (JSPS) KAKENHI Grant Number JP16K17660. M.\,E. and W.\,D.\,C. were supported by NASA grant NNX16AJ11G to The University of Texas. A.\,A.\,T. acknowledges support from JSPS KAKENHI Grant Number 17F17764. N.\,N. acknowledges support from KAKENHI Grant Number JP18H01265. A.\,P.\,H.,  Sz.\,Cs.,  S.\,G.,  J.\,K.,  M.\,P.,  and H.\,R.  acknowledge  support  by  DFG  grants  HA  3279/12-1, PA525/18-1,  PA525/19-1,  PA525/20-1,  and  RA  714/14-1 within the DFG Schwerpunkt SPP 1992, ``Exploring the Diversity  of  Extrasolar  Planets.'' The simulations were run on the Calculation Server at the NAOJ Center for Computational Astrophysics. This paper includes data collected by the \kepler mission. Funding for the \kepler mission is provided by the NASA Science Mission directorate. This work has made use of data from the European Space Agency (ESA) mission {\it Gaia} (\url{https://www.cosmos.esa.int/gaia}), processed by the {\it Gaia} Data Processing and Analysis Consortium (DPAC, \url{https://www.cosmos.esa.int/web/gaia/dpac/consortium}). Funding for the DPAC has been provided by national institutions, in particular the institutions participating in the {\it Gaia} Multilateral Agreement. This publication makes use of VOSA, developed under the Spanish Virtual Observatory project supported from the Spanish MINECO through grant AyA2017-84089. A.\,A.\,T. would like to thank Hori Yasunori and Michiko Fujii for helpful discussions on the system's dynamics.




\bibliographystyle{mnras}
\bibliography{references}







\bsp	
\label{lastpage}
\end{document}